# Exploitation of the nonresonant background of Multiplex-Coherent anti-Stokes Raman Scattering for label-free discrimination of proteins

M. Nafa[1], T. Mansuryan[1], V. Couderc[1], A. Tonello[1], L. Rechignat[1], M. Fabert[1], L. Magnol[2], V. Blanquet[2], F. Baraige[2], C. Carrion[3], J.-R. Duclère[4], C. Lefort[1*]

1. Université de Limoges, XLIM, UMR CNRS 7252, Limoges, France
2. Université de Limoges, Epidemiology of chronic diseases in tropical zone EpiMaCT, Inserm U1094, IRD U270, INRAE USC1501, Limoges, France
3. Université de Limoges, BISCEm, Microscopy core Facility, Limoges, France
4. Université de Limoges, IRCER, UMR CNRS 7315, Limoges, France

*claire.lefort@cnrs.fr

## Abstract

We propose a novel approach using Multiplex-Coherent Anti-Stokes Raman Scattering (M-CARS) for label-free discriminations in biomedical tissues. The strategy is based on the evaluation of the contrast between resonant and nonresonant contributions in a M-CARS hyperspectral dataset, and tested to identify and differentiate thin actin filaments from thick myosin filaments in muscle tissue without any labeling. First step consists in ensuring knowledge of the spatial regions containing thick myosin filaments thanks to its endogenous second harmonic signal, deducing expected location for thin actin filaments between myosin filaments. The ratio of resonant and nonresonant contributions for each pixel of the hyperspectral image allows then to discriminate actin from myosin filaments, whose localization is in accordance with the SHG probing. This qualitative imaging represents a proof of principle for highlighting and discriminating purposes in biological microscopy, thanks to the difference in the nonlinear properties of the related proteins. This paves the way for considering label-free imaging through a competition between two third-order nonlinear signatures.

## Introduction

Observation of biological tissues through light-matter interaction has been a hot research topic for decades. Label-free imaging strategies and diagnostics have continuously evolved technically due to the major challenge of eliminating the need for labeling biological substances, particularly in the case of in vivo imaging. To this end, label-free imaging strategies rely on the electrical susceptibility $\chi$ of the substance probed. A first family of approaches rely on the refractive index contrast ($\chi^{(1)}$) with techniques such as Differential Interference Contrast microscopy (DIC, also known as Nomarski microscopy) [1], tomography and Optical Diffraction Tomography (ODT) [2]. Their limitation concern samples that are thick and colored such as some biological tissues (DIC), and/or the lack of chemically specific response (ODT). For this purpose, the use of ultrafast laser pulses in the near infrared range enables to probe higher orders of the electrical susceptibility $\chi$. This permits specific imaging methods with excellent signal-to-noise ratio through nonlinear light-matter interaction resulting from the nonlinear electrical susceptibility of the substance probed. Multiphoton processes, such as two-photon fluorescence (TPF) or second harmonic generation (SHG), have demonstrated their ability to reveal several proteins label free. Indeed, in some favorable conditions, proteins can be autofluorescent, like elastine or keratine [3–6]. SHG is noteworthy, as a second order nonlinear susceptibility interaction ($\chi^{(2)}$), allowing for observation of non-centrosymmetric structures of proteins such as collagen or myosin networks [7,8].

Moreover, vibrational information can be revealed through Raman spectroscopy [9,10]. Coherent anti-Stokes Raman Scattering (CARS) [11] is of major interest in label-free imaging, as it exploits the third-order nonlinear optical susceptibility $\chi^{(3)}$. The signal contains a resonant part ($\chi^{(3)}_{R}(\Delta\omega)$) specific to given molecular groups contained in the matter across the Raman Shift ($\Delta\omega$), and a constant nonresonant background ($\chi^{(3)}_{NR}$) [9,12,13]. The latter is an electronic four-wave mixing process that interferes with the resonant response and alters the overall collected $|\chi^{(3)}(\Delta\omega)|$ spectrum [9,12].

Several techniques have been designed to reduce this nonresonant background. One way consists in acting on the polarization states of the pump, Stokes and probe beams [14], or acting on the delay between the creation of the ro-vibrational signature and its probing [14]. Another way consists in a post-processing reconstruction of resonant and nonresonant contributions through Maximum Entropy Methods (MEM), Kramers-Krönig transformations [15], to insist then on the sole resonant contribution. However, regarding multiplex CARS spectroscopy, Mueller et al. have not only used the nonresonant background to improve precision in concentration measurements [16], and but also suggested that nonresonant background may be beneficial in improving CARS resonant imaging [17].

Therefore, we propose to investigate the role of the nonresonant background in our M-CARS setup, consecutively to earlier works conducted by the team on both the metrology and the mapping of the second and third order nonlinear susceptibilities in inorganic structures [18,19]. Imaging was possible since, for such structures, $\chi^{(3)}_{NR}$ was proportional to the nonlinear refractive index $n_2$ [19]. In this publication, we measure the $\chi^{(3)}_{NR}/|\chi^{(3)}_{R}|$ ratio from the M-CARS hyperspectral image of a well-known mouse muscle model longitudinally excised, composed by alternating networks of actin and myosin filaments [20]. If myosin filaments constitute a non-centrosymmetric structure, which can therefore be characterized by SHG [7,21], actin filaments could not be efficiently discriminated without any labeling since both filaments share very similar Raman signatures [22,23]. The $\chi^{(3)}_{NR}$ /|$\chi^{(3)}_{R}$| ratio from the M-CARS hyperspectral image permits its label-free revelation.

## Experimental setup

The multiplex CARS (M-CARS) experimental scheme is presented in Figure 1 (a). The pump pulse used for our experiment is generated by a Q-switched microchip laser source (Horus Laser, 1064 nm, 1 ns, 20 kHz). Part of the pump beam is initially used to generate throughout a PCF fiber a supercontinuum signal, later used as Stokes signal. Its spectrum, acquired by means of an Optical Spectral Analyzer (OSA), is represented in Figure 1 (b) and is accompanied by its polynomial fit in the [520, 3300] $cm^{-1}$ Raman Shift range, corresponding to the spectral distribution of the M-CARS excitation. The rest of the pump signal undergoes through a half-waveplate followed by a Glan polarizer. The purpose is double: monitoring the pump energy on the one hand, and eliminating any cross-polarized contribution on the other hand. In our experiment, the excited vibrational modes are read by the probe beam, which also corresponds to the pump beam. Indeed, part of the nanosecond pulse serves in the creation of the ro-vibrational process, part of it to its reading.

Both Stokes and pump are respectively randomly (from the supercontinuum generation) and linearly polarized, so as to remain as neutral as possible regarding nonresonant background reduction or enhancement [14]. Once generated and spectrally filtered, the Stokes and pump beams are gathered in a collinear configuration by means of a dichroic mirror (Semrock, NFD01-1064-25x36), and are focused at the sample by a microscope objective (Obj 1: Olympus, UPlanSApo 60x, N.A. = 1.2, water immersion). The anti-Stokes signal is then generated among this interaction volume, and propagates collinearly with the other beams. In order to detect a collimated anti-Stokes signal, a second microscope objective (Obj 2: Nikon, S Plan Fluor ELWD 60x, N.A. = 0.7) is positioned to collimate the transmitted signal. A filter (Thorlabs, NF1064-44) is placed to remove any non anti-Stokes contribution before the detection system composed by the spectrometer (Horiba LabRAM HR Evolution, 600 gr/mm) and the CCD camera (Synapse CCD camera, actual spectral resolution < 0.8 $cm^{-1}$). As an example, a 2D image of curcumin crystals is also shown in Figure 1 (b) merging several C=O vibrational signatures at 1350, 1480 and 1630 $cm^{-1}$. A M-CARS trace obtained on a paraffin sample is shown in Figure 1 (d) revealing the $CH_2$ twist, bend, and stretching vibrational signatures with increasing Raman Shifts, and therefore demonstrating the effectiveness of our system.

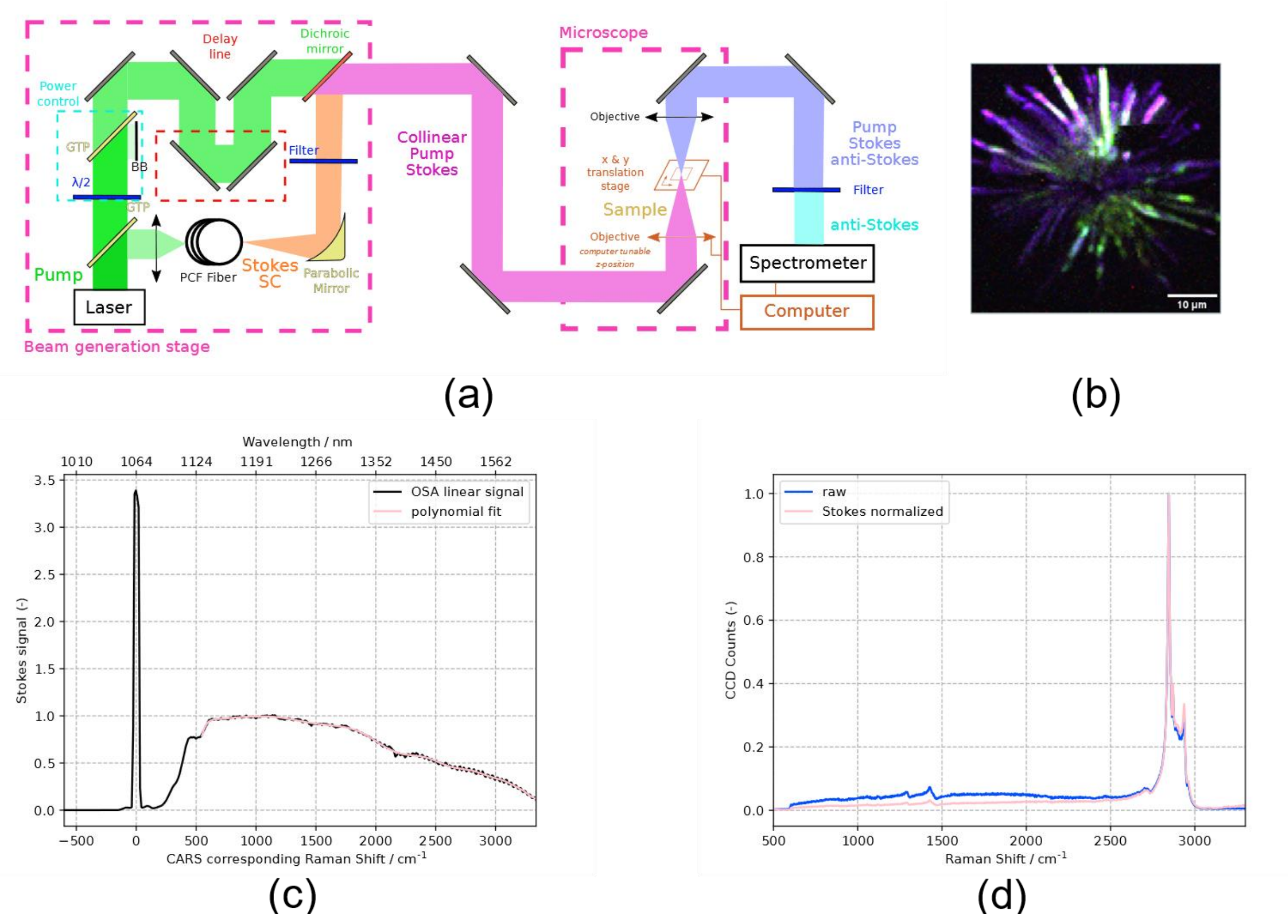

*Figure 1: (a) M-CARS experimental setup scheme, from beam generation stage to anti-Stokes detection, through CARS interaction at the sample. λ/2, half-wave plate; GTP, Glan-Taylor polarizer; PCF, photonic crystal fiber; BB, beam blocker; SC, supercontinuum. (b) An example of curcumin superparticle composite image is also presented. (c) Spectrum of the Stokes supercontinuum pulse (black) acquired on an Optical Spectrum Analyzer (OSA). The abscissa has been adapted in corresponding CARS Raman Shift scale, taking as reference our narrow pump pulse at 1064 nm. A polynomial fit has been performed on the operating range of the M-CARS setup and is represented in pink. (d) Presentation of paraffin broadband M-CARS spectra: raw (blue); normalized by Stokes supercontinuum profile (pink). CH2 twist, bend, and stretching vibrational signatures are noticeable.*

As test sample, we have chosen the mouse muscle model with a 4 µm thick longitudinally excised Extensor Digitorum Longus (EDL) muscle. Its ultrafine structure is made of sarcomeres, representing the elementary contractile unit, composed by an alternative assembly of thin filaments of actin and thick filaments of myosin [6,15]. This sample is placed on a microscope slide (model Superfrost), without any embedding medium.

The optical focus is first realized in bright-field conditions near the edge of the muscle, leading to the muscle image shown in Figure 2 (a). The latter contains several labels. First, there is a yellow box representing the scanning area for each hyperspectral imaging presented image presented later on. Besides, two colored markers show where typical M-CARS spectra are acquired on the muscle (white) but also on the microscope slide to define a baseline signal (orange). These representative spectra, presented in Figure 2 (b), are the topic of the next section and have a double objective: estimating the resonant and nonresonant contributions of the M-CARS signal; defining a relevant baseline signal from the microscope slide that will be further subtracted to better highlight biological M-CARS signals.

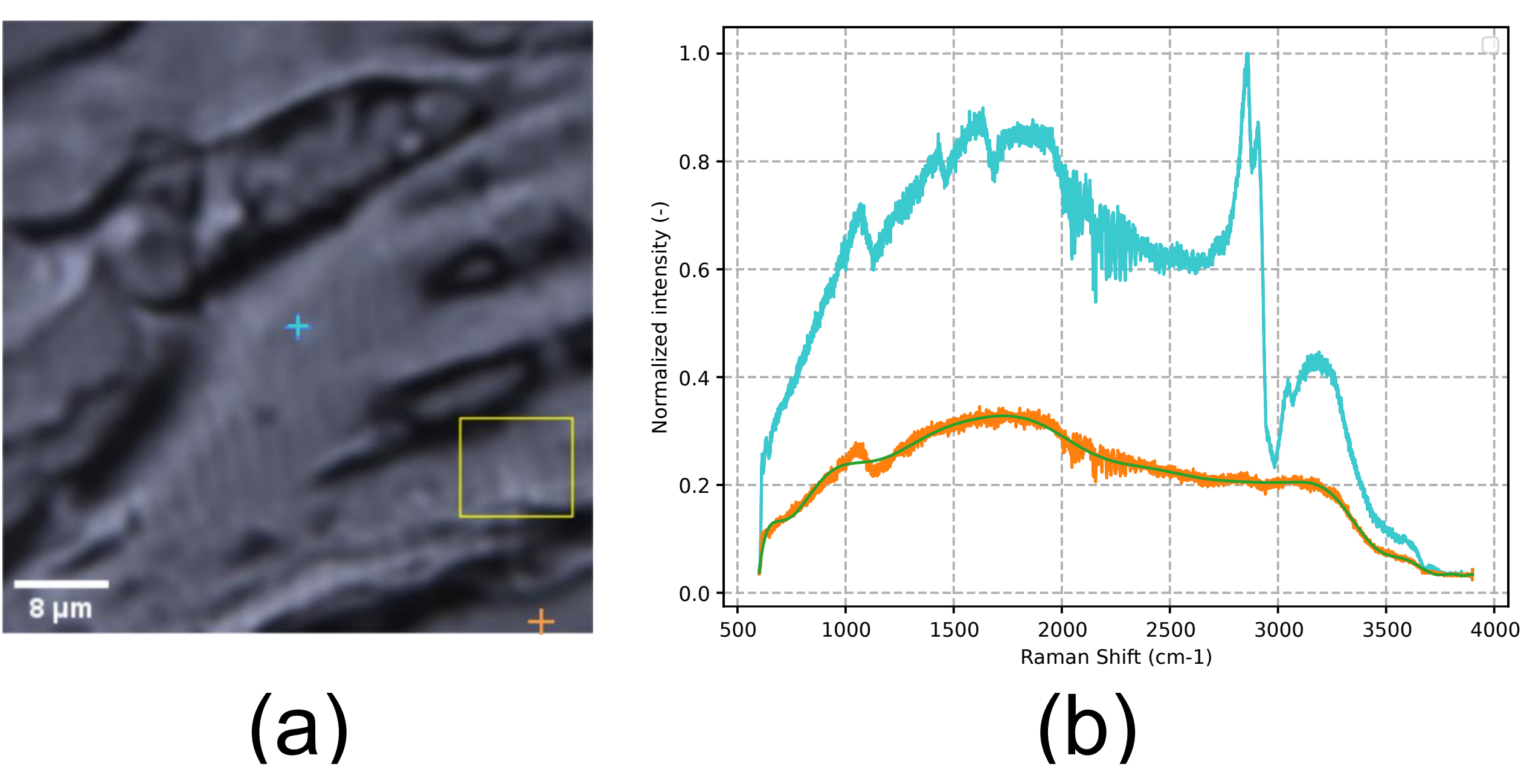

*Figure 2: Imaging from the muscle sample: (a) Bright-Field illumination; (b) Broadband acquisition of CARS spectra at the markers spotted in (a). The orange spectrum is considered as the response of the Superfrost microscope slide, which polynomial fit is also drawn in green.*

## Identification of resonant and nonresonant M-CARS contributions

Each broadband M-CARS spectrum shown in Figure 2 (b) is acquired over a 1s exposure time. The orange spectrum, acquired outside the muscle, reveals the M-CARS signal of the microscope slide alone. The usefulness of this spectrum is to establish a reference baseline, fitted through a polynomial function represented in green (Figure 2 (b)), compared to which the M-CARS spectrum of the muscle, in cyan, can be fully interpreted. Expected features can be observed in the latter. Weak signals, expected in a protein structure as muscle [24], can be identified in the fingerprint region, respectively the -CH bend around 1450 $cm^{-1}$, and the amide I contribution around 1650 $cm^{-1}$. A strong resonant CARS response can be observed around 2900 $cm^{-1}$ corresponding to the $-CH_2$ and $-CH_3$ stretching groups [5,25], where the resonant contribution $\chi^{(3)}_R$ is predominant in the $|\chi^{(3)}(\omega)|$ response. We can also see around 3000 $cm^{-1}$ a minimum in intensity followed by a rounded signal. This rounded signal corresponds to the symmetric vibration feature of -OH[25], but is also reshaped due to interference of the CARS signal with the nonresonant background ($\chi^{(3)}_{NR}$). Therefore, we identify the Raman shift span [2600, 3300] $cm^{-1}$ as our region of interest, containing not only the $|\chi^{(3)}_R|$ information (between 2820 and 2920 $cm^{-1}$) but also the $\chi^{(3)}_{NR}$ information (further from resonance, for example in the 2620-2670 $cm^{-1}$ range). Concerning the glass baseline subtraction, we proceed as follows. For each pixel of the M-CARS hyperspectral mapping, the polynomial fit of in Figure 2 (b) is normalized and subtracted so that the dip value around 3000 $cm^{-1}$ reaches a value close to zero. The structuring idea is finally to obtain a M-CARS signal, having a resonant and nonresonant contribution both exclusively originating from the muscle.

# Results and discussion

## Resonant M-CARS hyperspectral imaging

M-CARS hyperspectral image has been recorded by mapping the region highlighted in yellow in Figure 2 (b), where the glass baseline signal is subtracted as previously explained. For each spectrum of the hyperspectral image, the integration time is 0.1 s, for a total acquisition time of 36 minutes. Namely, the M-CARS resonant signal is considered with the $-CH_2$ and $-CH_3$ stretching contribution ($\chi^{(3)}_R$) between 2820 and 2920 $cm^{-1}$, which integration imaging is represented in Figure 3 (a). On this picture, the presence of thick filaments can be guessed, with a poor contrast respect to its blurry neighboring areas. Away from resonance, between 2620 and 2670 $cm^{-1}$, the integration leads similarly to an even blurrier image (Figure 3 (b)). This confirms that on these separate spectral ranges of detection, the differences between actin and myosin regions are indistinguishable, and that no clear information on actin structure can be detected, as expected [19]. Therefore, we exploit the SHG signal from our pump pulse as an endogenous marker for the location of thick myosin filaments, as presented in the next subsection.

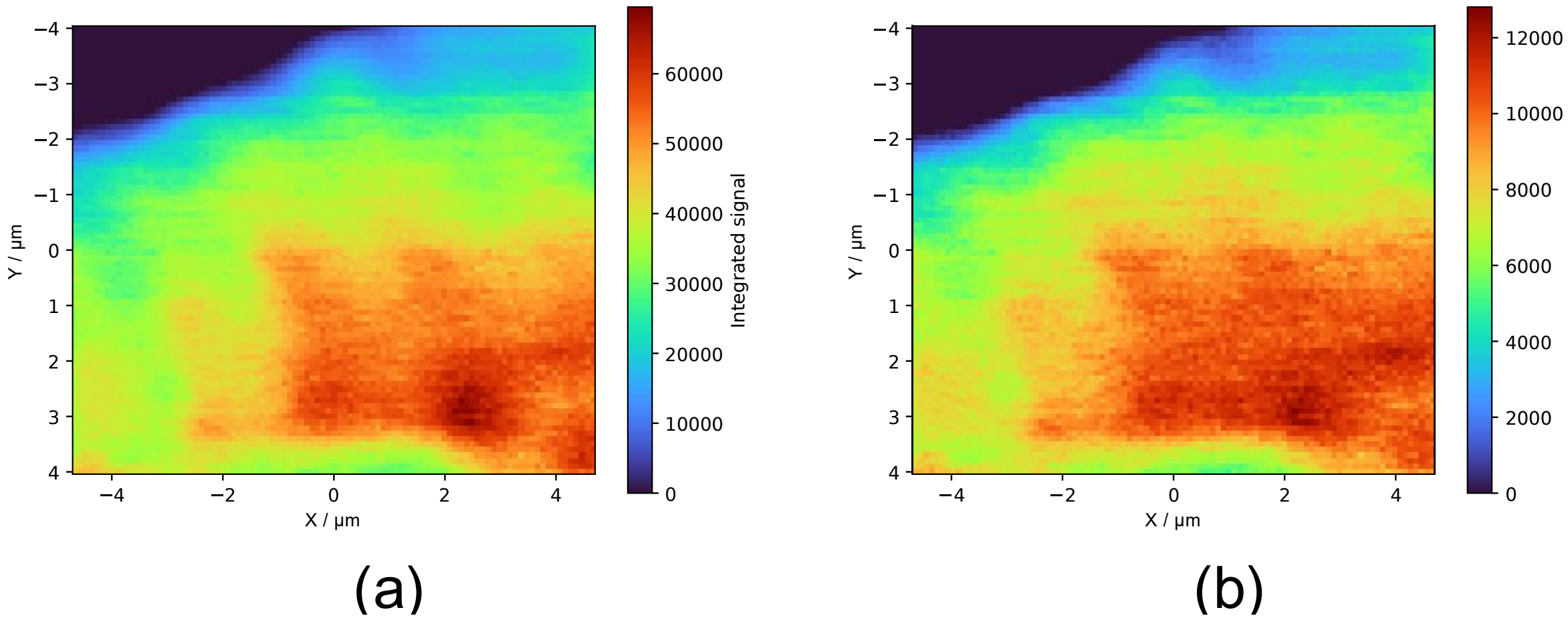

*Figure 3: M-CARS hyperspectral imaging integrated between: (a) 2820 and 2920 cm-1 ; (b) 2620 and 2670 cm-1*

## Identification of myosin structure through SHG imaging

In order to identify the second harmonic generation from thick myosin filaments, the detection wavelength range is centered at 532 nm, leading to the image shown in Figure 4 (a). The SHG signal reveals the presence of myosin filaments which are spaced by around 2 µm, corresponding to the usual spacing between them [6, 16]. Localization of thin actin filaments can then be presupposed, since they are known to be contained between myosin structures identified by SHG. Therefore, by considering the M-CARS hyperspectral datacube (x, y, Raman Shift), two groups of M-CARS spectra can specifically be considered, on the basis of their (x,y) location, and being attributed to regions containing thick myosin or *a priori* thin actin filaments. The aim of the next subsection is to highlight average differences in the M-CARS spectra for such filaments.

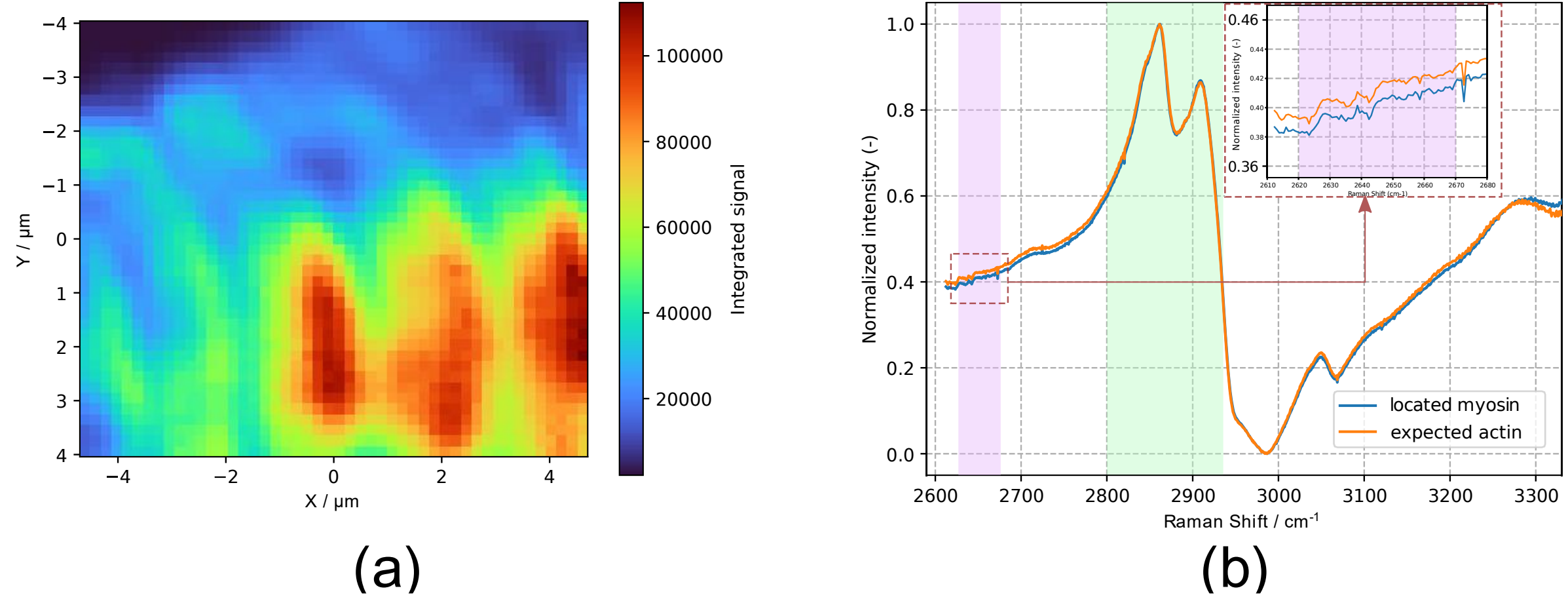

*Figure 4: (a) Second Harmonic Generation (texp = 0.1s per pixel) image revealing thick myosin filaments.(b) Average M-CARS spectra from locations highlighted as myosin (blue) or actin (orange). Colored stripes represent Raman Shift regions where nonresonant contribution (pink) and resonant contribution (green) are predominant.*

## Average M-CARS study on actin and myosin filaments

For each group, we processed the mean M-CARS spectrum (respectively 1440 and 2430 spectra for actin and myosin were considered) presented in Figure 4 (b), normalized first by the Stokes energy profile distribution of Figure 1 (b), and then to its maximal value. The robustness of the baseline subtraction can be appreciated since the minimum signal always stands near zero. Here, the M-CARS resonant signal is considered with the $-CH_2$ and $-CH_3$ stretching contribution ($|\chi^{(3)}_R|$), i.e. between 2820 and 2920 $cm^{-1}$ (green stripe). In this range, the spectral difference between actin and myosin regions is negligible, as previously shown in Figure 3. However, valuable information can still be extracted from Figure 4 (b). We can notice that further away from the resonance, in the pink stripe where the nonresonant contribution ($\chi^{(3)}_{NR}$) is predominant, the normalized signal in Figure 4 (b) stands around 0.4, and that the difference between spectra is almost constant, around 0.025. This is an interesting result since it could explain the slight differences observed on the integrated images at and further from resonance, respectively, in Figure 3 (a) and in Figure 3 (b), although they were unable to reveal any actin structure. The 2620-2670 $cm^{-1}$ Raman Shift range in **Erreur ! Source du renvoi introuvable.** (b) can then be used as a $\chi^{(3)}_{NR}$ reference for a discriminative $\chi^{(3)}_{NR} / |\chi^{(3)} (\Delta\omega = 2860\ cm^{-1})|$ ratio approach, where $|\chi^{(3)} (\Delta\omega = 2860\ cm^{-1})|$ is hereafter designed as $|\chi^{(3)}_R|$ due the pre-dominance of the resonant contribution for this Raman Shift value. This also means that this spectral range can be used to estimate the relative nonlinear properties of these biological structures, as was done for inorganic structures [19]. The next subsection corresponds to the application of this strategy of using the $\chi^{(3)}_{NR} /|\chi^{(3)}_{R}|$ ratio for each singular M-CARS spectrum of the hyperspectral datacube.

## Nonresonant contribution enhancement for actin filament discrimination

For each M-CARS spectrum of the hyperspectral dataset, the maximal value has been assigned to the CARS resonant contribution $|\chi^{(3)}_{R}|$ between 2820 and 2920 $cm^{-1}$, and the $\chi^{(3)}_{NR}$ value has been estimated as the mean value between 2620 and 2670 $cm^{-1}$. The corresponding $\chi^{(3)}_{NR} /|\chi^{(3)}_{R}|$ images are presented in Figure 5 (a) and (b) for the respective ratio ranges 0.41-0.48 and 0.36-0.41 corresponding to their colorbars. It is noticeable that the highlighted features in Figure 5 (a) now correspond to what lies between the thick filaments of myosin, that can be identified in **Erreur ! Source du renvoi introuvable.** (b). A 1D plot is performed in Figure 5 (c) along the white lines depicted in Figure 5 (a) and Figure 5 (b). We notice that the maxima of the first image correspond to the minima of the other image and vice versa, as expected in muscle between actin and myosin. The revelation of thin actin filaments is confirmed by its trimmed superposition made with the SHG image in Figure 5 (d). The 1D plot performed on the latter, shown in Figure 5 (e), illustrates that, once again, the maxima of the myosin image correspond to the minima of the other image. Therefore, we claim that this signal represents the thin actin filaments that are located between the thick filaments of myosin. This result demonstrates that thin actin filaments can now be specifically visualized label-free with a correct signal-to-noise ratio, without any addition of fluorophore thanks to its difference in nonlinearities with thick myosin filaments.

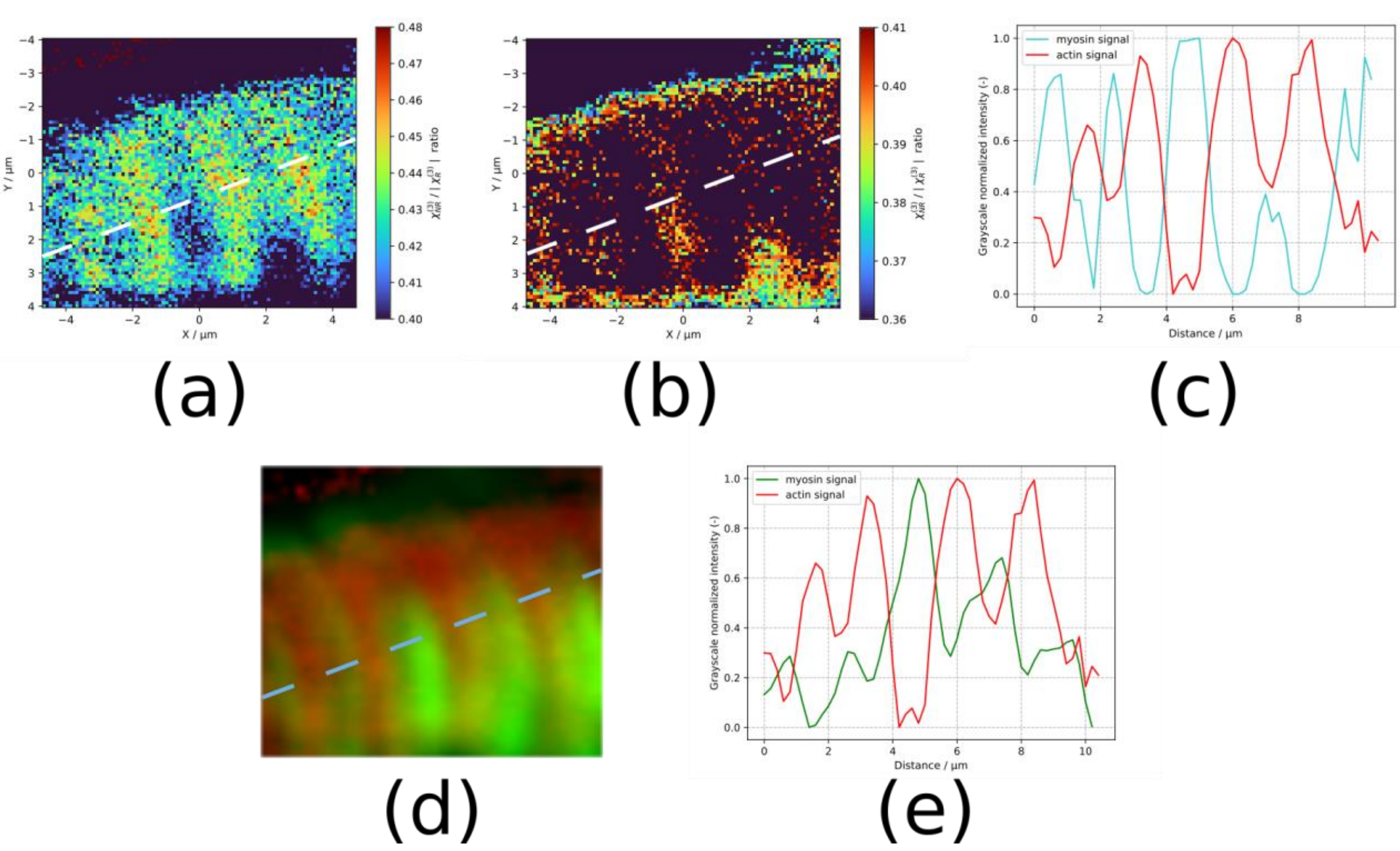

*Figure 5: $\chi^{(3)}_{NR}$ /|$\chi^{(3)}_{R}$| ratio obtained from M-CARS hyperspectral mapping exploiting the interaction between resonant and nonresonant background, the latter being considered as the mean value between 2620 and 2670 cm-1. Highlighted regions are: (a) actin ; (b) myosin. The dashed white lines indicates the direction of the (c) plotted 1D-cropped signals. (d) Superposition of the actin image (red) with the SHG image indubitably highlighting thick myosin filaments (green, extracted from Figure 4 (a)). The dashed blue line indicates the direction of the (e) plotted 1D-cropped signals.*

Also, how this image has been obtained, due to the $\chi^{(3)}_{NR}$ /|$\chi^{(3)}_{R}$| ratio, is particularly interesting for several reasons. The first is practical: performing a ratio is a simple task that is compatible with real-time observations. The second is that the label-free discrimination of the fine protein structures of actin and myosin filaments is made possible by measuring their difference in nonlinear properties. The fact that we keep dividing by the maximal resonant response on each pixel enables to circumvent various instabilities of the interaction volume created by the crossing of the pump and Stokes beams on the biological sample. The third reason is that we considered the interaction between several four-wave mixing processes on the same spectrum, by exploiting the nonresonant background. This gives additional information than standard strategies which commonly consider several light-matter interaction processes one by one in multimodal imaging. One also has to consider that a broadband

excitation and a sufficient spectral resolution for the probe pulse are necessary to be able to consider not only resonant CARS features, but also Raman Shifts far away enough that carries nonresonant CARS contribution.

## Conclusion

In this paper, a M-CARS setup utilizes a supercontinuum Stokes pulse generated from a narrowband pump pulse to enable multimodal hyperspectral imaging. A M-CARS imaging technique, based on the $\chi^{(3)}{}_{NR}$ /$|\chi^{(3)}{}_{R}|$ ratio, enables label-free imaging of actin thin filaments with a satisfactory signal-to-noise ratio. This qualitative imaging is a proof-of-principle that nonresonant signal can be helpful in discriminating and identifying biological structures, confirmed on the well-known case of thin actin filaments located in-between thick myosin filaments. More generally we can state that label-free imaging can not only be considered singularly through one light-matter interaction effect, but rather through the competition involving several of them. The perspectives of this work are multiple. A first one can be to determine the optimal polarization state for the supercontinuum Stokes pulse so that nonresonant interaction is enhanced. Another one could be to discriminate other biological substances of interest.

## Acknowledgments

This work was partially funded by the LabEx Sigma-lim ANR-10- LBX-0074, laboratory of excellence launched by the French Ministry of Higher Education and Research (https://www.unilim.fr/labex-sigma-lim) between IRCER (https://www.ircer.fr) and the XLIM (https://www.xlim.fr) Research Institutes. We thanks the MITI from CNRS which have financially supported the Multikhi project.